\documentstyle[pre,twocolumn,aps,eqsecnum,epsfig]{revtex}
\def\e{{\rm e}}
\begin{document}

\twocolumn[\hsize\textwidth\columnwidth\hsize\csname
@twocolumnfalse\endcsname

\title{Magneto-Transport in the Two-Dimensional Lorentz Gas}
\author{A. Kuzmany and H. Spohn}

\address{Theoretische Physik, Ludwig-Maximilians-Universit\"at, 
Theresienstr. 37,
D-80333 M\"unchen, Germany\\
email: spohn@stat.physik.uni-muenchen.de, 
       kuzmany@stat.physik.uni-muenchen.de
}

\date{submitted to Physical Review E, September 30, 1997}

\maketitle
\begin{abstract}
We consider the two-dimensional Lorentz gas with Poisson distributed
hard disk scatterers and a constant magnetic field perpendicular to
the plane of motion. The velocity autocorrelation is computed
numerically over the full range of densities and magnetic fields with
particular attention to the percolation threshold between hopping transport and pure edge currents. The Ohmic and
Hall conductance are compared with mode-coupling theory and a recent
generalized kinetic equation valid for low densities and small fields. We
argue that the long time tail as $t^{-2}$ persists for non-zero
magnetic field. 
\\

\noindent
PACS number(s): 05.60.+w, 05.20.Dd, 75.50.Jt
\end{abstract}
\vskip2pc]
%\pacs{05.60.+w,05.20.Dd,75.50.Jt}
\section{Introduction}
Two-dimensional electron films can be manufactured with high
perfection in GaAs
heterostructures. At low temperatures a mean free path of over $10^4nm$ is
reached and to a very good approximation the electrons may be
considered as noninteracting. To have some interesting physics one 
nanostructures the probe by lithographic or other techniques. Thereby a 
strongly repulsive potential is imposed
on the electrons with a maximum above the Fermi energy ({\em quantum
antidots}). 
If the imprinted structure is on a scale larger than the Fermi wave
length, adjustable to be of the order of $50nm$, one hopes to 
capture the transport properties already in the
classical approximation. 

So far, the most popular geometry has been a
regular array of antidots which corresponds classically to the Sinai
billard and should result quantum mechanically in the
Hofstadter butterfly. The magneto-transport of this
periodic structure has been studied in great detail, both
experimentally \cite{1,2,3,4} and theoretically \cite{5,6,7,8}. In
our note we investigate randomly placed antidots. To our knowledge, 
the so far best experimental realization has been achieved by L\"utjering \cite{9}. 
We compare our results with his measurements in the conclusions. 

In kinetic theory randomly distributed scatterers are known as the Lorentz gas, which has
proved to be an important testing ground. In particular, one can
understand precisely the assumptions for the validity of the (linear) Boltzmann
equation \cite{10} and check on the accuracy of the low density
expansion and its non-analytic character \cite{11,12,13,14,15,16}. 
Also the long time tails in the velocity autocorrelation function
are seen most convincingly in
the Lorentz gas \cite{17}. In distinction from the work mentioned we investigate here 
the dynamics in presence of a magnetic field perpendicular to the plane of
motion. 

The model has a strong geometric flavour: one places randomly disks
(the scatterers) of radius $a$ in the plane at density $n_s$. The
disks may overlap. In the region outside the disks we have independent
point particles with density $n_e$. They have mass $m^*$, charge $e$, and
move in a uniform external magnetic field $B_{ex}$. Thus a single
particle travels along a circle and is elastically reflected upon
collision with a scatterer. We denote the velocity of the particle at
time $t$ by ${\bf u}(t)$. Clearly $|{\bf u}(t)|$ is conserved and we
set it equal to the Fermi velocity $v_F$, since at low temperatures
contributions to the transport only come from the Fermi surface.
The radius of gyration is then $R_e = v_Fm^*/eB_{ex}$.

We are interested in the magneto-transport which relates the steady
state current ${\bf j}$ to an in plane uniform electric field 
${\bf E}$ by
\begin{equation}\label{1.1}
{\bf j}=\sigma {\bf E}
\end{equation}
for small $\bf E$ and want to understand how $\sigma$ depends on
$B_{ex}$ and $n_s$.
$\sigma_{11}$ and
$\sigma_{22}$ are the Ohmic conductivities. 
In our case $\sigma_{11}=\sigma_{22}$ by
isotropy. $\sigma_{12}=-\sigma_{21}$ is the Hall conductivity.
The magneto-transport will be studied in linear response. This means
the dynamics is the one just explained ({\em zero} electric field)
and the transport coefficients are given in terms of the
time-integrated velocity autocorrelation functions. 
Physically one has to average over all of phase
space. For a large sample, this is equivalent to fixing the initial
position and averaging over the scatterer distribution. It is this
prescription we will use both theoretically and in the
numerics. Before spelling out the details we introduce dimensionless
quantities. 

Space is measured in units of the disk radius $a$ and velocity in
units of the Fermi velocity $v_F$. 
The scatterers have then radius
$1$ and their dimensionless density is $\rho=n_s a^2$. The 
dimensionless radius
of gyration is $R=R_e/a$ and the corresponding magnetic field is
$B=1/R= eaB_{ex}/v_F m^*$. 
Let ${\bf v}(t) = {\bf u}(t)/v_F$ be the velocity of the
particle at time $t$ starting at the origin. Clearly  $|{\bf
v}(t)|=1$. ${\bf v}(t)$ depends on ${\bf v}(0)$ and the
particular configuration of the scatterers. 

We define the dimensionless velocity autocorrelation function by
\begin{equation}\label{1.2}
C_{ij}(t)=\langle v_i(0)v_j(t)\rangle\;\;,i,j=1,2\,.
\end{equation}
Here $\langle\;\cdot\;\rangle$ is a double average.  
Firstly there is an average over scatterers. The centers of the
disks are distributed according to a Poisson process with uniform
density $\rho$ conditioned on the set $\{{\bf x}\mid |{\bf x}|\leq 1\}$
being free of centers. Secondly we average over the initial
velocity ${\bf v}(0)=(\cos\varphi, \sin\varphi)$ uniformly in
$\varphi$. (By rotational invariance this second average could be
omitted, but it is of advantage numerically.) The conductivity 
tensor is then  
\begin{equation}\label{1.3}
\sigma_{ij}=\frac{n_e e^2}{ m^*}\,D_{ij}\,,\;
D_{ij}=\int \limits_0^\infty dt\,\langle v_i(0)v_j(t)\rangle\,.
\end{equation}
$D$ depends on $\rho$ and $B$. 
Also of interest is the frequency dependent conductivity defined by
\begin{equation}\label{1.4}
\sigma_{ij}(\omega)=\frac{n_e e^2}{ m^*}\,D_{ij}(\omega)\,,\;
D_{ij}(\omega)=\int \limits_0^\infty dt\,\e^{i\omega t}
\langle v_i(0)v_j(t)\rangle\,.
\end{equation}
We note that $\langle v_i(0)v_j(t) \rangle = \langle v_i(0)v_j(-t) \rangle$
by stationarity 
and $\langle v_1(0)v_2(t) \rangle = - \langle v_1(0)v_2(-t) \rangle =
- \langle v_2(0)v_1(t) \rangle$ by time reversal. 
Therefore $D_{12} = - D_{21}$ and 
$D_{ii}=(1/4)\,\,\int_{-\infty}^\infty dt\,\langle{\bf v}(0)\cdot
{\bf v}(t)\rangle$.

As stated in (\ref{1.3}) the conductivity is ill-defined. There is 
always a non-zero probability that the particle will not
be scattered at all. If so, $\langle
v_1(0)v_1(t)\rangle = (1/2)\; \cos(Bt)$ and $\langle
v_1(0)v_2(t)\rangle = (1/2) \;\sin(Bt)$, and the time integral in (\ref{1.3}) 
needs a
reinterpretation. Physically there will always be a weak elastic
scattering by impurities, i.e. once in a while the velocity direction
is randomized. In approximation the velocity autocorrelation function
is then
modified to 
\begin{equation}\label{1.5}
\e^{-t/\tau} \langle v_i(0)v_j(t)\rangle
\end{equation}
and the proper definition reads
\begin{equation}\label{1.6}
%\sigma_{ij}(\omega)=\frac{e^2 m^*}{\pi \hbar^2}\,D_{ij}(\omega)\,,\;
D_{ij}(\omega)= \lim_{\tau \rightarrow \infty}
\int \limits_0^\infty dt\,\e^{-t/\tau}\e^{i\omega t}
\langle v_i(0)v_j(t)\rangle\,.
\end{equation}
At zero density, no scattering, (\ref{1.6}) results in 
\begin{equation}\label{1.7}
D_{11}(\omega)=-i\,\frac{\omega}{2}\,\frac{1}{B^2-\omega^2}\,,\:
D_{12}(\omega)=\frac{B}{2}\,\frac{1}{B^2-\omega^2}\,,
\end{equation}
as well known from the Drude theory of the Hall effect. In particular, 
at $\omega=0$ both the Ohmic conductivity and the Ohmic resistance vanish. 

It is instructive to go back for a moment to the case of an in plane 
electric field $\bf E$. One finds that 
for freely moving particles the average
current approaches 
\begin{equation}\label{1.8}
{\bf j}=\frac{1}{2B}\,(-E_2,E_1)
\end{equation}
as $t\rightarrow\infty$ in accordance with (\ref{1.7}) at $\omega=0$. In fact, one
would expect that scattering cannot make things worse. Thus for the
Lorentz gas there should be a well-defined steady state current $\bf
j$ for $B\neq 0$. We did not find a general argument to establish its
existence. The situation at $B=0$ is very different. 
Then the particle is accelerated along the direction of $\bf E$. 
Since only the 
velocity direction is randomized by collisions, the energy input from the 
electric field cannot be dissipated and no meaningful steady state current is 
reached as $t\rightarrow\infty$. However for small 
$\bf E$ the time dependent current settles at a 
plateau over a time span the longer 
the smaller $\bf E$, whose value equals $\sigma {\bf E}$ with $\sigma$
of (\ref{1.3}). 

To give a brief outline: In the following section we discuss the dependence of 
$D$ on $\rho$, $B$ with particular attention 
to the two percolation thresholds. We also compare our numerics with the mode 
coupling theory of G\"otze and Leutheu{\ss}er \cite{18}, 
which seems to be the only theoretical prediction at intermediate densities. 
Recently Bobylev et al. \cite{19} derived a 
generalized transport equation for low scatterer densities and small fields. 
In the appropriate domain of validity their 
predictions are in fact very accurate (Section 3), and clearly improve on the 
phenomenological Boltzmann equation with magnetic 
field. One of the famous results on the Lorentz gas at $B=0$ is the slow decay 
of the velocity autocorrelation function as $-t^{-2}$ for large $t$ in two 
dimensions \cite{20}. For densities $\rho<0.25$ such a power law has been well 
established numerically \cite{17}. 
At larger densities there is a pre-asymptotic decay 
approximately as $-t^{-1.4}$ 
and, with reasonable numerical effort, the true 
asymptotics cannot be seen any more. In Section 4 we argue that also for $B\neq0$ 
the velocity autocorrelation 
decays as $t^{-2}$ for large $t$. 
We conclude by a comparison with the experiments of L\"utjering and with some comments. 

\section{The static conductivity}
To discuss $D$ in its dependence on $\rho$, $B$ at zero frequency it is useful 
to consider first the limiting cases. As explained $D_{11}(0,B)=0$, 
$D_{12}(0,B)=1/2B$ at $\rho = 0$. On the 
other hand for $B=0$ we have $D_{12}(\rho,0)=0$. 
$D_{11}(\rho,0)$ has been studied numerically \cite{16,17}. For $\rho
\rightarrow 0$ one obtains the Boltzmann value $D_{11}(\rho)=3/16\rho$. 
Thus close 
to $(\rho,B)=0$, $D$ is somewhat singular and roughly of the form 
$(\rho^2+B^2)^{-1/2}$. Its precise functional dependence will be discussed in 
Section 3. 

As the density is increased the disks percolate. This means that for 
$\rho>\rho_c$, $\rho_c\cong 0.36$ with probability one the origin is contained 
in a finite domain bounded by scatterers. 
For $\rho<\rho_c$ the origin is connected to infinity by a path not 
intersecting 
the scatterers. Close to $\rho_c$ numerically one finds
$D_{11}(\rho,0)\cong 
|\rho-\rho_c|^{1.5}$, $\rho \leq \rho_c$, with no theoretical explanation yet. 

In fact, the Lorentz gas has a second percolation threshold. Let us fix  
$\rho<\rho_c$ and increase $B$. Above some critical value $B_c$, the 
trajectory 
will either be a circle or skip along a , possibly large, cluster of 
a finite number of disks. The bulk current for $B<B_c$ is reduced to a pure edge current. 
Since the particle cannot leave a 
cluster, the mean square displacement is bounded which 
implies $D_{11}(\rho,B)=0$
for $B>B_c$ by the Einstein relation
$2D_{11}=\lim_{t\rightarrow\infty}
\langle {\bf r}^2(t)\rangle/2t$. 
A hopping of the particle from cluster to cluster is only possible, if disks 
with radius $1+R$ percolate which means, 
in our units, $\rho(1+R)^2\geq \rho_c$, i.e. 
\begin{equation}
\label{2.1}
B_c=\frac{1}{\sqrt{\rho_c/\rho}-1}\,.
\end{equation}
Strictly speaking the $B_c$ of (\ref{2.1}) is only an upper bound on the true 
$B_c$. 
One could imagine that already for a slightly smaller $B$ hopping is
suppressed. Numerically, we see a smooth variation through $B_c$ 
and such a fine
point cannot be decided. In Fig. 1 we plot the two 
domains in which the Ohmic
conductivity vanishes.

The behavior of $D_{12}$ is somewhat more complicated. We always have the
contribution of the circle orbits. According to the Poisson distribution
their probability is $\exp[-\pi\rho\kappa(R)]$ with $\kappa(R)=R(R+2)$ 
for $R<1$ and
$\kappa(R)=4R-1$ for $R\geq 1$. We decompose the average correspondingly as
\begin{equation}\label{2.2}
\langle v_i(0)v_j(t)\rangle_{\geq 1}+\langle v_i(0)v_j(t)\rangle_0\,,
\end{equation}
where $\langle\;\cdot \;\rangle_{\geq 1}$ is the average over all initial ${\bf v}(0)$
and all scatterer configurations such that there is at least one collision for
$0\leq t<\infty$. The normalization is
$\langle 1 \rangle_1=1-\exp[-\pi\rho\kappa(R)]$. If $\langle
v_i(0)v_j(t)\rangle_{\geq 1}$
is absolutely integrable, then
\begin{equation}
D_{ij}=\lim\limits_{t\rightarrow\infty}
\langle v_i(0)[x_j(t)-x_j(0)]\rangle_{\geq 1}
+(1-\delta_{ij})\frac{R}{2}\e^{-\pi\rho\kappa(R)}\,,
\label{2.3}
\end{equation}
where ${\bf x}(t)=\int_0^tds\,{\bf v}(s)$ is the position of the particle at
time $t$. The average $\langle v_i(0)[x_j(t)-x_j(0)]\rangle_{\geq 1}$ can be further 
decomposed into averages counting
with how many different disks the particle collides in the course of time. The
lowest contribution corresponds to the particle skipping around a single disk,
etc..
Now, if either $B>B_c$ or $\rho>\rho_c$, then by assumption the sum over all
clusters with a {\em finite} number of disks exhausts already the complete
average $\langle v_i(0)[x_j(t)-x_j(0)]\rangle_{\geq 1}$ for arbitrary $t$. 
(In the domain where $D_{11}>0$, 
the particle collides with an
infinite number of disks). Let us fix then a particular finite cluster of
disks and let $\Gamma$ be the set $\{{\bf x},{\bf v}\}$ of initial 
conditions such that the
particle will eventually collide with each one of the disks in the cluster
and no others. By assumption $|{\bf x}|\leq const.$ 
for $\{{\bf x},{\bf v}\}\in \Gamma$. 
If the dynamics
in $\Gamma$ is mixing (or decomposes into mixing components), then
$\lim_{t \rightarrow\infty}\langle v_1(0) [x_2(t)-x_2(0)]\rangle_\Gamma=
\langle v_1(0)\rangle_\Gamma\langle  x_2(0)\rangle_\Gamma
-\langle v_1(0) x_2(0)\rangle_\Gamma=-\langle v_1(0) x_2(0)\rangle_\Gamma$, 
since $\langle v_1(0) \rangle_\Gamma = \langle \frac{d}{dt} x_1(t) 
\rangle_\Gamma = 0$ by stationarity. 

For one disk the motion is integrable \cite{21} and the mixing 
assumption fails.
In the Appendix we compute the contribution $D_{12}^{(1)}$, to (\ref{2.3}) coming from rosette orbits around a single disk. 
The case of two non-overlapping
disks was studied in \cite{21}. They still found elliptic islands, however
with a small measure already. 
The mixing assumption seems to be satisfied approximately. One might hope that the remainig contribution, $-\langle v_1(0) x_2(0)\rangle_\Gamma$, integrated over allowed scatterer configurations averages to zero. Numerically this does not seem to be the case. At density $\rho=0.1$, the circling and rosette orbits together account only for $50\%$ to the observed $D_{12}$ in the range $1.5\le B\le2.5$, cf. Fig. 2b.

To have a more complete picture of $D$ we simulate the Lorentz gas
numerically. For given $B$ and scatterer configuration we compute
$v_j(t)$ up to $60$ collision times. The system size is chosen so large that
${\bf x}(t)=\int_0^t ds\,{\bf v}(s)$ never hits the boundary. 
To speed up the simulation we use a
hierarchical search for the next point of collision. For each scatterer
configuration we average over $100$ randomly chosen initial velocity 
directions. For
$C_{ij}(t)$ to be sufficiently smooth typically one has to average then over
$10^6$ sample paths.
The conductivity is determined from (\ref{2.3}). In most of parameter space
$\langle v_i(0) x_j(t)\rangle$ has not yet reached its 
asymptotic value, which reflects the
slow decay of the velocity autocorrelation functions. 
We essentially extrapolate ``by hand'' to $t\rightarrow\infty$ which results in a 
slight overestimate whenever there is
an independent check. 
In Figs. 2-5
we display our results at 
densities $\rho=0.1,\,0.15,\,0.2$ and $0.3$. The percolation
threshold is indicated by a vertical line. Note that for
$\rho=0.1$ the $B$-scale starts at $B=0.5$.
Our data show a fairly smooth interpolation of the asymptotics at $B=0$, $B=\infty$. 
The most
surprising feature is an initial increase of the Ohmic conductivity with $B$ at
intermediate densities. Apparently, the curved trajectory can bend
itself more easily through the
dense ``labyrinth'' of scatterers. The Hall conductuctivity rises steeply to its
maximum and then levels off. 
The maximum is shifted to smaller $B$ as the density decreases.
In Fig. 2b we insert the contribution of the circling orbits according to (\ref{2.3}) and of rosette orbits according to (\ref{A.9}). 

To our knowledge the only attempt to derive the 
magneto-conductivity at intermediate densities is
the mode coupling theory of G\"otze and Leutheu{\ss}er \cite{18}. In
Fig. 6 we compare their prediction
with our simulation data, note the particular choice of units. 
For the lowest density, $\rho=0.032$, we find good
agreement. Essentially the same behaviour is obtained from the generalized
Boltzmann equation, to be discussed in 
the following section. However, the
kinetic theory fails already at density $\rho=0.13$ whereas 
mode coupling is still a
reasonable approximation. The next higher density $\rho=0.25$ can no longer be
accounted for.

\section{Low Density, Small Fields}
For $B=0$ the linear Boltzmann equation is exact at low density. More
precisely, for $\rho\rightarrow 0$ collisions of the particle with the same
scatterer become unlikely and the density of particles in phase space on the
scale of the mean free path is governed by the Boltzmann equation. In
particular this yields $D_{11}(\rho,0)=3/16\rho$ as $\rho\rightarrow 0$.
With external magnetic field $B$ one generalizes in the obvious way to
\begin{eqnarray}\label{3.1}
\partial_t f({\bf
x},\varphi,t)=(-\cos\varphi\;\partial_1-\sin\varphi\;\partial_2-B\,\partial_\varphi)
f({\bf x},\varphi,t)
\nonumber
\\
+2\rho\int\limits_{-\pi}^\pi
d\varphi'\,\frac{1}{4}\sin|\frac{\varphi'}{2}|[f({\bf
x},\varphi-\varphi',t)-f({\bf x},\varphi,t)]\,.
\end{eqnarray}
Here $f$ is the distribution function at ${\bf x}$, 
\\
${\bf
v}=(\cos\varphi,\sin\varphi)$, $t$.
On the basis of (\ref{3.1}) one obtains
\begin{equation}
\label{3.2}
D_{11}^0=\frac{1}{2}\,\frac{8\rho/3}{(8\rho/3)^2+B^2}\,,\;D_{12}^0=\frac{
1}{2}\,\frac{B}{(8\rho/3)^2+B^2}\,,
\end{equation}
which we rewrite in scaling form as
\begin{eqnarray}\label{3.3}
D_{11}^0=\frac{1}{\sqrt{\rho^2+B^2}}\,g_{11}^0(\rho/B)\,,\;
\nonumber
\\
D_{12}^0=\frac{1}{\sqrt{\rho^2+B^2}}\,g_{12}^0(\rho/B)\,,\;
\nonumber
\\
g_{11}^0(y)=\frac{1}{2}\,\sqrt{1+y^2}\,\frac{8y/3}{(8y/3)^2+1}\,,\;
\nonumber
\\
g_{12}^0(y)=\frac{1}{2}\,\sqrt{1+y^2}\,\frac{1}{(8y/3)^2+1}\,.\;
\end{eqnarray}

For a rigorous derivation the radius of gyration must be of the order of the
mean free path. Thus $B$ must vanish linearly with $\rho$. 
As observed by Bobylev et al. \cite{19} even in the limit
$\rho\rightarrow 0$ some recollisions survive. This is most easily seen for
circle orbits with no collisions at all. According to the Boltzmann equation
(\ref{3.1}) even after several turns the particle still has some 
probability to be
scattered. However, for the mechanical Lorentz 
gas after one completed revolution the
annulus is surely free of disks and no scattering events 
can occur. In \cite{19} the circling orbits and
the recollisions are properly taken into account 
for $\rho\rightarrow 0$ and a
generalized kinetic equation with memory term is derived. On this basis the
velocity autocorrelations are computed. The conductivity is still of 
the scaling form (\ref{3.3}),
\begin{eqnarray}\label{3.4}
D_{11}^*=\frac{1}{\sqrt{\rho^2+B^2}}\,g_{11}^*(\rho/B)\,,\;
\nonumber
\\
D_{12}^*=\frac{1}{\sqrt{\rho^2+B^2}}\,g_{12}^*(\rho/B)\,,\;
\end{eqnarray}
which is just a consequence of $B/\rho = const.$ as $\rho \rightarrow
0$.. However the scaling functions are now
modified to 
\begin{equation}\label{3.5}
g_{11}^*(y)=\frac{1}{2}\,\sqrt{1+y^2}\,\frac{1}{(2y\gamma(x))^2+1}
(1-x^2)2y\gamma(x)\,,\;
\end{equation}
\begin{equation}\label{3.6}
g_{12}^*(y)=\frac{1}{2}\,\sqrt{1+y^2}\,\frac{1}{(2y\gamma(x))^2+1}
(1+x^2(2y\gamma(x))^2)\,,\;
\end{equation}
\begin{equation}\label{3.7}
x=\e^{-2\pi y}\,,
\end{equation}
\begin{equation}\label{3.8}
\gamma(x)=1-\frac{1-x^2}{2x^2}
\left(\frac{1-x^2}{2x}\ln\frac{1+x}{1-x}-1\right)\,.
\end{equation}
In Fig. 7 we plot $g_{11}^*$, $g_{12}^*$. 
Note that in terms of the polar 
angle in the $\rho$-$B$-plane the scale is highly compressed at the
left. 
In Fig. 8 we
plot the correction to the Boltzmann value (\ref{3.3}). 

Of course, $D^*$ reproduces the correct limiting behaviour 
for $\rho\rightarrow 0$ 
and $B\rightarrow 0$. According to (\ref{2.1}) the 
percolation boundary behaves as
$B_c\cong 1.66\sqrt\rho$ for small $\rho$. Since for the 
validity of (\ref{3.4}) 
$B$ is scaled proportional to $\rho$, $D^*$ cannot see this threshold. 

In Figs. 9, 10 we compare our numerical 
results with $D^*$. We also include 
$D^0$. As expected, $D^*$ is a considerable improvement. Note that 
the agreement
is not uniform in $\sqrt{B^2+\rho^2}$. This reflects that 
at $B=0$ the Boltzmann 
equation has a restricted range of validity, e.g. 
$D_{11}(\rho,0)/D_{11}^0(\rho)=0.65$ at $\rho=0.1$. 

\section{velocity autocorrelations}
On the level of the linear Boltzmann equation (\ref{3.1}) one obtains
an exponential decay for the velocity autocorrelations. Explicitely
\begin{equation}
C_{11}^0(t)=\frac{1}{2}\e^{-|t|/\tau_0}\cos(Bt)\,,\;
C_{12}^0(t)=\frac{1}{2}\e^{-|t|/\tau_0}\sin(Bt)\,,
\end{equation}
with $\tau_0=(8\rho/3)^{-1}$. The correct low density/ small field
behaviour has a more interesting structure. To state the result it is
convenient to introduce the Laplace
transforms
\begin{equation}
F(z)=\int\limits_0^\infty dt\,\e^{-zt}2(C_{11}(t)+iC_{12}(t))\label{4.2}
\end{equation}
for $\Re e (z)>0$. 
In the same approximation as leading to (\ref{3.4}) one obtains 

\begin{equation}
F(z)=\frac{1-\e^{-(z+\nu)T}}{z-i\omega+\nu
\int\limits_{-\pi}^{\pi}d\psi\,g(\psi)
\frac{1-\e^{i\psi}}{1-\e^{(z+\nu)T +i\psi}}}+
\frac{\e^{-(z+\nu)T}}{z-i\omega}\,.
\label{4.3}
\end{equation}
We find it convenient to compare the Fourier transforms 
\begin{equation}
\hat{C}_{ij}(\omega)=
\int\limits_{-\infty}^\infty dt\,\e^{-i\omega t} C_{ij}(t)
\end{equation}
with the
simulation data. Since $C_{11}(t)$ is even and $C_{12}(t)$ is odd, we have
$\hat{C}_{11}(\omega)=\Re e
(F(i \omega)+F(-i \omega))$, 
$\hat{C}_{12}(\omega)=i\Re e (-F(i \omega)+F(-i \omega))$. 
The circle orbits yield $\delta$-peaks at $\omega=
\pm 1/R$ with weight $\exp(-4\pi \rho R)$. In Fig. 11 we
plot the prediction from (\ref{4.3}) and compare it with the
numerics.
Note that the $\delta$-peaks are out of scale. As anticipated the
agreement is excellent, in fact over the whole low density/ small
field regime. We plot $\hat{C}_{12}(\omega)$, which turns out to be negative,
Fig. 12. As the ratio $\rho/B$ is increased, the
characteristic double peak merges into a single peak which shifts then
to $\omega=0$.

Away from low density/ small fields we have no
theory to compare with. For $B=0$ $C_{11}(t)$ has the long time tail
of the form $C_{11}(t)=-\alpha t^{-2}$ with $\alpha>0$. We refer to
\cite{18} for a detailed discussion. The heuristic explanation runs as
follows: At long times the main contribution of $C_{11}(t)$ comes from
paths returning to the origin at time $t$. Let ${\bf v}(0)=(1,0)$. Then the
particle will be more likely to return from the right which yields a
negative correlation in the velocity. For the excursion away from the
origin we use a random walk approximation. If at some intermediate
time the particle arrives at the line $x=0$, then it will return to
the origin equally likely from right and left and the contribution to
$C_{11}(t)$ vanishes. Therefore the tail can be computed from a return
to the origin at time $t$ of a random walker {\em without} ever
hitting the line $x=0$. This probability decays as $t^{-2}$ in two
dimensions. Clearly, in our argument we only used that the motion
before returning to the origin is diffusive. This remains valid for
non-zero $B$. Thus $C_{11}(t)$ and $C_{12}(t)$ should have a decay as
$t^{-2}$ for $t\rightarrow\infty$. 
For $B=0$ the long time tail is most clearly seen at $\rho=0.15$ \cite{17}. 
We increase $B$ to $0.2$ and average over $3\cdot 10^7$ sample paths. 
The effective exponent for both correlations is $1.9$ with a negative prefactor, 
Fig. 13. However the prefactor of $C_{11}(t)$ becomes positive in the 
range $B=0.5\,...\,0.8$ and of $C_{12}(t)$ in the range 
$B=0.6\,...\,3$. Thus when the particle returns to the origin it picks up
more complicated velocity correlations than in the case $B=0$. As one 
example of such a sign reversal in the $1,2$-correlation we display the 
data for $\rho=0.2$, $B=2.3$, Fig. 14. The effective exponents are approximately $1.4$.

\section{discussion}
We studied transport in the two-dimensional Lorentz gas with a
constant magnetic field perpendicular to the plane of motion. The
theory of Bobylev et al. is in fair agreement with our simulation
data, both for the transport coefficients and the velocity
autocorrelations. Away from low density/small magnetic fields there is
little theory to compare with. The qualitative properties of the
magnetotransport can be guessed from the well understood limiting
cases $B\rightarrow 0$ and $B\rightarrow\infty$. The most unexpected
feature is an increase in the Ohmic conductivity with increasing $B$
at intermediate densities. 
Such a behavior has also been found experimentally \cite{9}. 

In these measurements antidots are imprinted at random locations with 
densities $1/(1000nm)^2$, $1/(600nm)^2$, $1/(400nm)^2$, $1/(300nm)^2$, 
and $1/(240nm)^2$, resp. . 
The electrons feel a screened potential. For the periodic case this is 
usually modelled by $V(x_1,x_2)=V_0|\sin (\pi x_1/a)\sin (\pi x_2/a)|^\beta$ 
with $\beta$ ranging from $2$ to $4$. For a random distribution the true 
potential is more complicated, in particular when there is strong overlap. 
Thus we cannot expect quantitative agreement between our hard disk model 
potential and the experiment \cite{9}. There the reduced density 
roughly ranges from $0.01$ to $0.25$ in our units.  
The dimensionless magnetic field varies from $0$ to $3$. Qualitatively the conductivities 
$D_{11}$, $D_{12}$ in dependence on $B$ follow our curves. However the measured 
$D_{12}$ is smaller by approximately a factor $1/2$. At the largest density a distinct increase 
in $D_{11}$ with $B$ close to $B=0$ is observed. As in our numerical studies the percolation threshold, $B_c$, is hardly seen in the Hall conductance. The Ohmic conductivity is small 
for $B>B_c$, but the transition region is fairly broad. 

The velocity autocorrelations have a slow decay over the full range 
of parameters. Presumably it is governed by $t^{-2}$, which is however 
severely masked by an even slower preasymptotic decay. The 
definite sign of the prefactor at $B=0$ is not retained. 

\acknowledgments
We thank W. Schirmacher for bringing the thesis of G. L\"utjering to our attention. 

\appendix
\section{Hall conductivity of rosette orbits}
We determine the contribution to $D_{12}$ from orbits which circle 
around a single scatterer, cf. Section 2. Since $B> B_c$ , we discuss only the case
$R< 1$. There is a corresponding formula for $R\geq 1$. 

We choose a large square box $\Lambda$. Let $y=\{{\bf y}_1,...,{\bf y}_n\}$ be the centers 
of the scatterers and $P_\rho (dy)$ their normalized Poisson 
distribution at density $\rho$. We set $\Lambda (y)=\{{\bf x}\in\Lambda\mid 
|{\bf x}-{\bf y}_j|\geq 1,\, j=1,...,n\}$ and denote 
by $|\Lambda(y)|$ the area of this set. Let $F_j(t;{\bf x},\varphi,y)=\cos\varphi
\sin\varphi(t;{\bf x},\varphi,y)$ if $\varphi(t)$ defines a rosette orbit at scatterer $j$ and
$F_j=0$ otherwise. Then
\begin{equation}
\label{A.1}
\langle v_1(0)v_2(t)\rangle_1=\int P_\rho(dy)\frac{1}{|\Lambda(y)|}\int\limits_
{\Lambda(y)}d^2x \frac{1}{2\pi} \int d\varphi\sum\limits_{j=1}^n F_j(t)\,.
\end{equation}
We have $|\Lambda(y)|\cong  |\Lambda| \e^{-\pi\rho}$. By translation symmetry we shift
the scatterer $j$ to the origin and introduce the polar coordinates ${\bf x}=(r\cos\alpha, r\sin\alpha)$. 
Let $f_r(t;\alpha, \varphi)=\cos\varphi\sin\varphi(t;r, \alpha, \varphi)$ if the orbit with initial 
center of gyration $(r\cos\alpha, r\sin\alpha)$ and initial velocity $(\cos\varphi,\sin\varphi)$ defines a rosette 
orbit around the scatterer located at $0$ and $f_r=0$ otherwise. Let $d(y)=
\min_j |{\bf y}_j|$. 
Taking the limit $|\Lambda|\rightarrow\infty$ we obtain 
\begin{eqnarray}
\label{A.2}
\nonumber
\langle v_1(0) v_2(t)\rangle_1=
\rho\e^{\pi\rho}\int P_\rho(dy)\chi\{2\le d(y)\}\times
\\
\nonumber
\int\limits_{1-R}^{(1+R)\wedge (d(y)-1-R)} dr \, r 
\frac{1}{2\pi}\int d\alpha\int d\varphi f_r(t,\alpha,\varphi)=
\\
\nonumber
\rho\e^{\pi\rho}\int P_\rho(dy)\chi\{r+R+1\le d(y)\}\times
\\
\nonumber
\int\limits_{1-R}^{1+R} dr \, r 
\frac{1}{2\pi}\int d\alpha\int d\varphi f_r(t,\alpha,\varphi)=
\\
\rho\e^{\pi\rho}
\int\limits_{1-R}^{1+R} dr \, r 
\e^{-\pi(r+R+1)^2\rho}
\frac{1}{2\pi}\int d\alpha\int d\varphi f_r(t,\alpha,\varphi)
\end{eqnarray}
with $\chi(A)$ denoting the characteristic function of the set $\{A\}$. 

As before, the Hall conductivity of rosette orbits is given by
\begin{equation}
\label{A.3}
D_{12}^{(1)}=
\lim\limits_{z\rightarrow 0}\int\limits_0^\infty dt\, \e^{-zt}
\langle v_1(0) v_2(t)\rangle_1\,.
\end{equation}
By rotation invariance, the $\alpha$ integration becomes trivial. Thus we have to determine
\begin{equation}
\label{A.4}
\lim\limits_{z\rightarrow 0}\int\limits_0^\infty dt\, \e^{-zt}
\int d\varphi f_r(t;\alpha,\varphi)\, .
\end{equation}
We fix $\alpha=\frac{3\pi}{2}$ and $r$ such, that $1-R\le r\le 1+R$. 
Then $-\varphi_+\le \varphi\le \varphi_+$, 
with 
\begin{equation}
\label{A.5}
\varphi_+=\arccos((1-R^2-r^2)/(2Rr))\, .
\end{equation}
$\varphi(t)$ increases linearly in $t$. At a collision $\varphi(t)$ jumps by the angle
\begin{equation}
\label{A.6}
\Psi=2\arccos((r^2-R^2-1)/2R)\, .
\end{equation}
Thus 
\begin{equation}
\label{A.7}
\varphi(t)=\varphi+\frac{t}{R}+n(t, \varphi)\Psi\,,
\end{equation}
where $n(t,\varphi)$ is the number of collisions up to time $t$ 
for the initial $\varphi(0)=\varphi$. Inserting in (\ref{A.4}) yields
\begin{eqnarray}
\label{A.8}
\nonumber
\lim\limits_{z\rightarrow 0}\int\limits_0^\infty &dt&\, \e^{-zt}\int\limits_{-\varphi_+}^{\varphi_+} d\varphi\,
\cos\varphi\sin(\varphi+\frac{t}{R}+n(t,\varphi)\Psi)=
\\
\nonumber
\lim\limits_{z\rightarrow 0} &R& \int\limits_{-\varphi_+}^{\varphi_+} 
d\varphi\,\cos\varphi
(2i(-zR+i))^{-1}\times
\\
\nonumber
&[&(e^{\varphi_+(-zR+i)}\e^{zR\varphi}-\e^{i\varphi})
+(\e^{2\varphi_+(-zR+i)}-1)\times
\\
\nonumber
\sum\limits_{n=0}^\infty&\e&^{i(n+1)\Psi}\e^{(2n+1)(-zR+i)\varphi_+}\e^{zR\varphi}]
+c.c.
=
\\
\nonumber
&R& \int\limits_{-\varphi_+}^{\varphi_+} d\varphi\,\cos\varphi
[(\cos\varphi-\cos\varphi_+)-
\\
\nonumber
&(&2(1-\cos(2\varphi_+ +\Psi)))^{-1}\times
\\
\nonumber
&(&\cos(3\varphi_+ +\Psi)-\cos(\varphi_++\Psi))]=
\\
\nonumber
&R&(\varphi_+-\frac{1}{2}\sin 2\varphi_+)+
\\
\nonumber
&2&R(\sin\varphi_+)^2\sin(2\varphi_++\Psi)(1-\cos(2\varphi_++\Psi))^{-1}\, .
\\
\end{eqnarray}
Combined with (\ref{A.2}) we obtain
\begin{eqnarray}
\label{A.9}
\nonumber
&D&_{12}^{(1)}=
R\int\limits_{1-R}^{1+R}dr\, r\rho \e^{\pi\rho(1-(r+R+1)^2)}
\{\varphi_+-\sin\varphi_+\cos\varphi_++
\\
&2&(\sin\varphi_+)^2\cos(\varphi_++(\Psi/2))/\sin(\varphi_++(\Psi/2))\} \, .
\nonumber
\\
\end{eqnarray}
In Fig. 2b we plot our result (\ref{A.9}) for $\rho=0.1$ as a function of 
$B=1/R$.

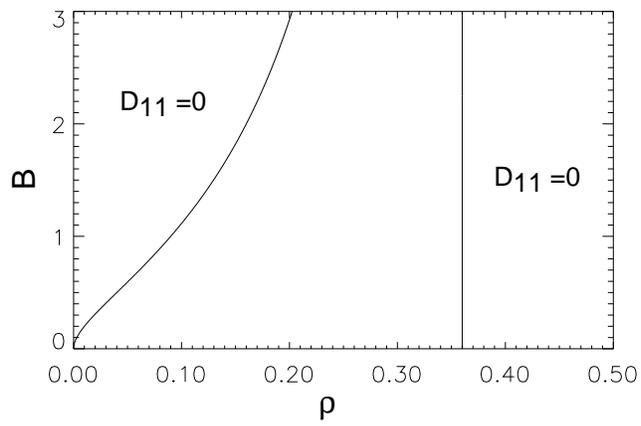
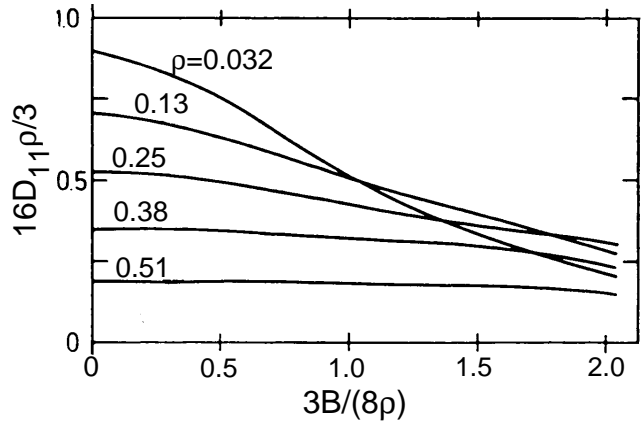
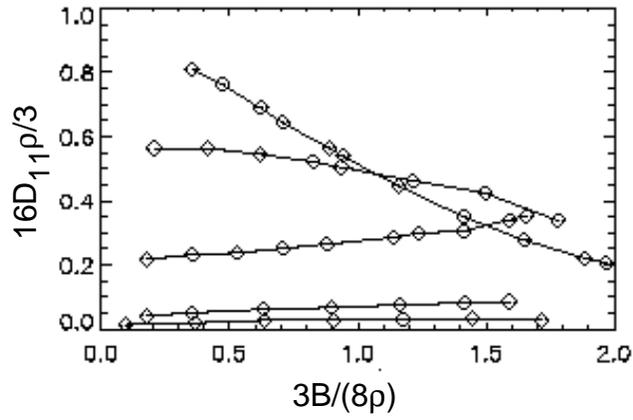
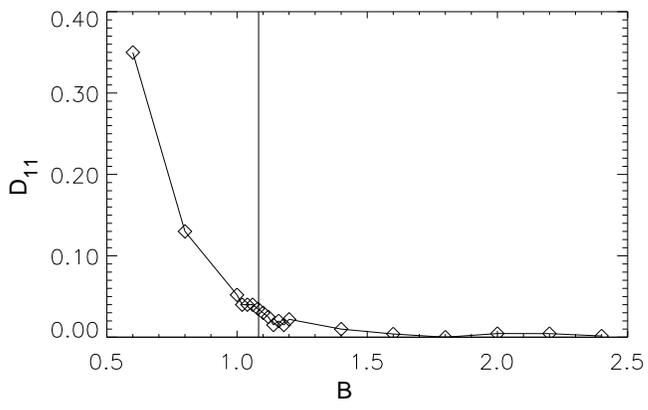
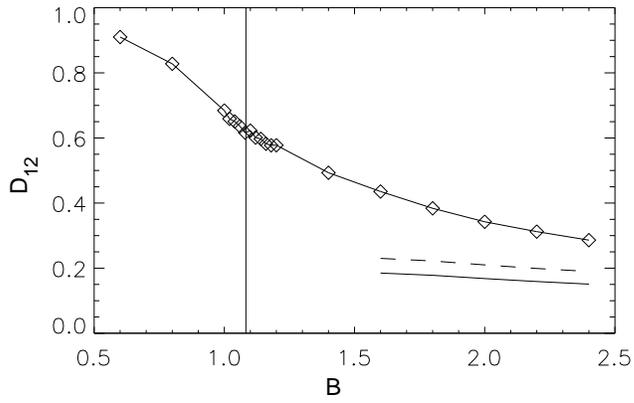
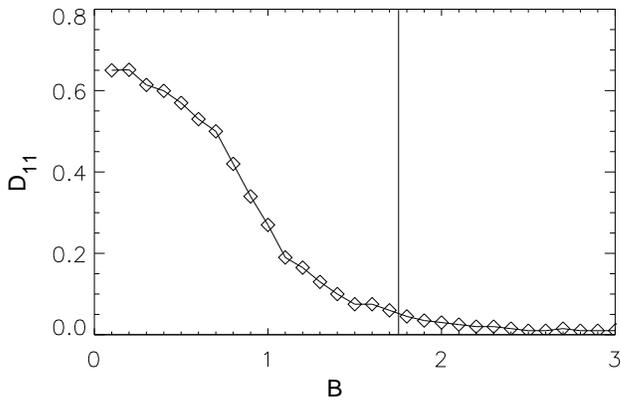
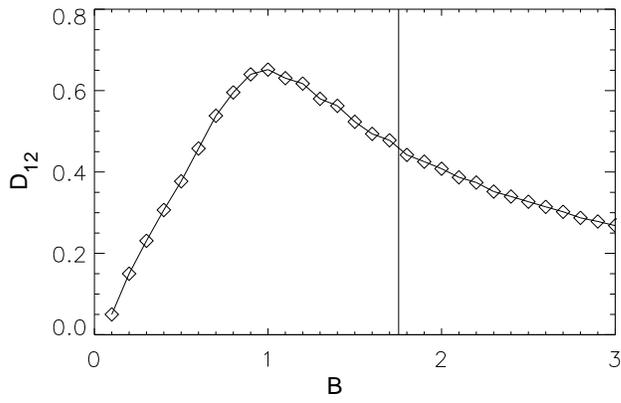
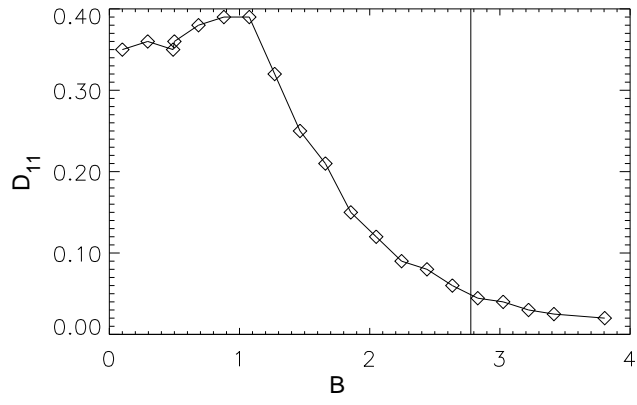
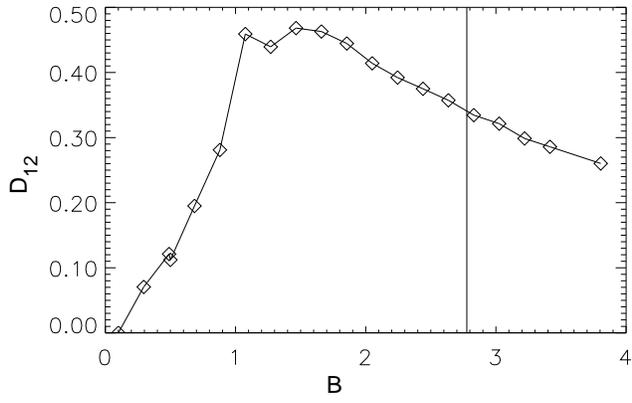
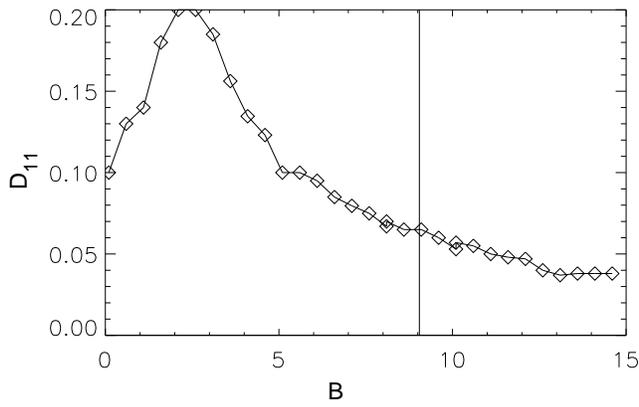
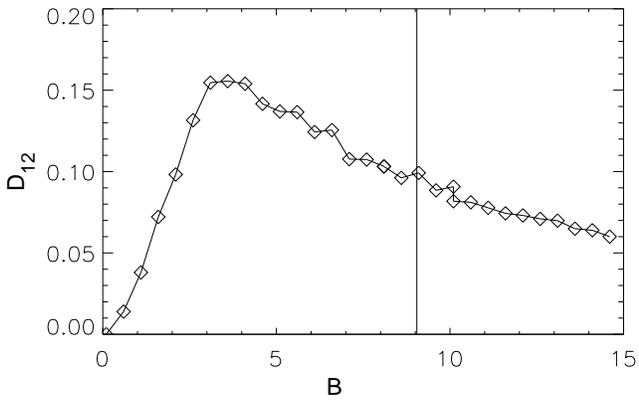
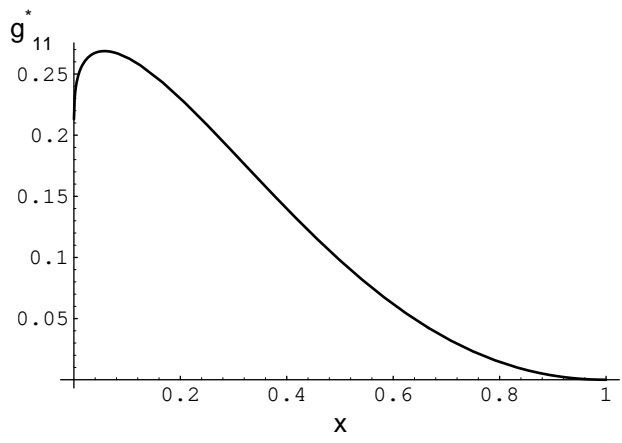
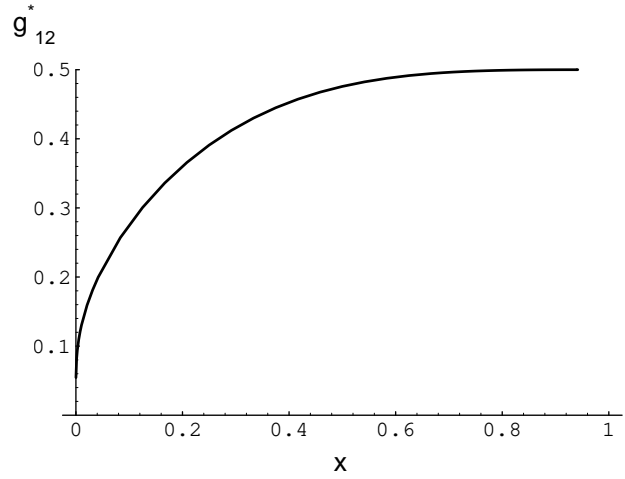
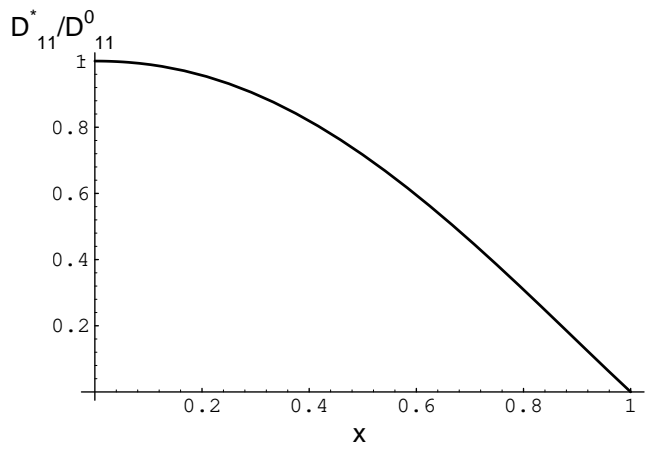
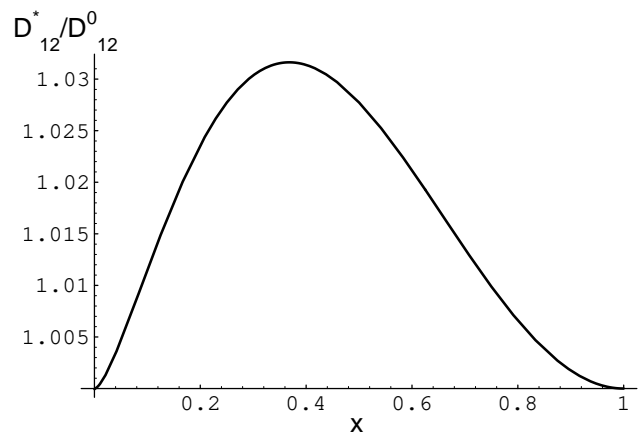
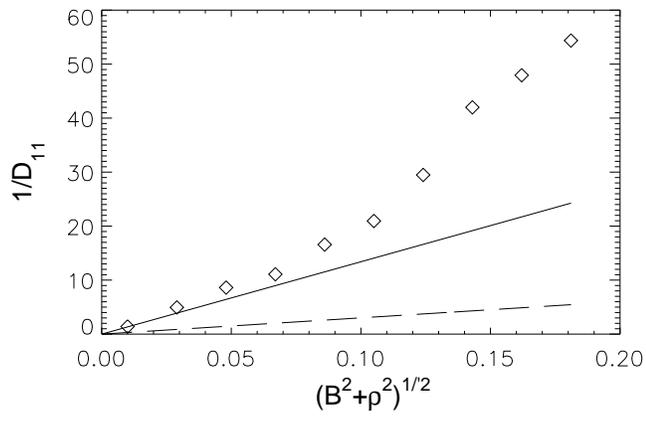
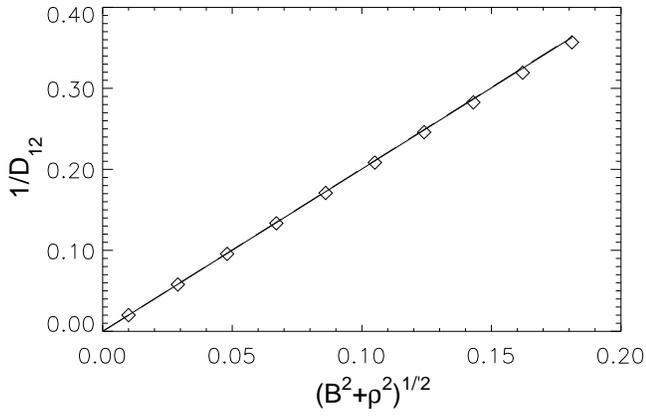
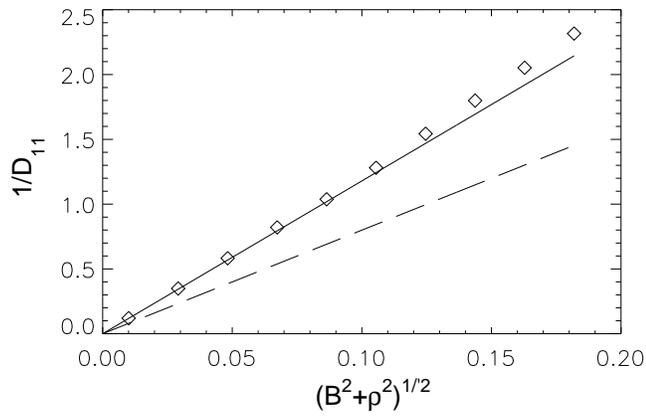
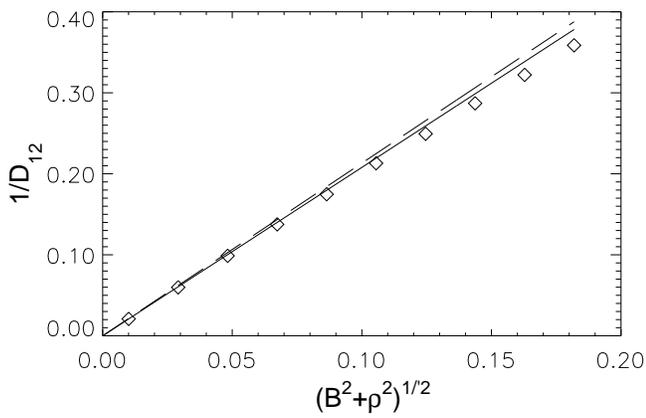
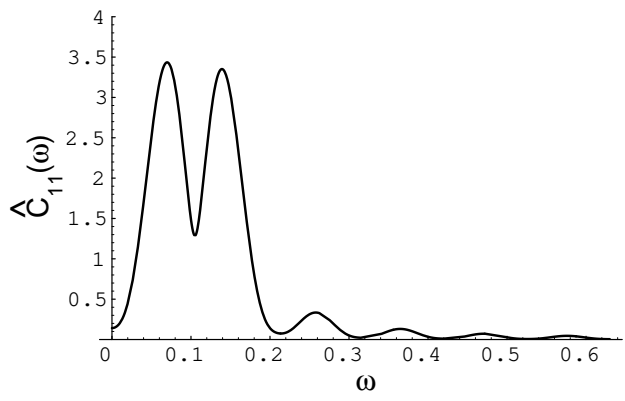
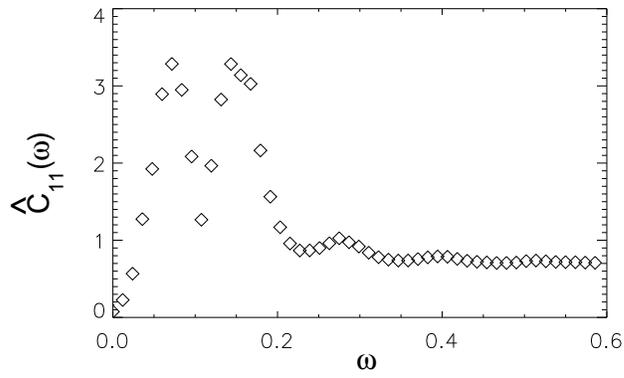
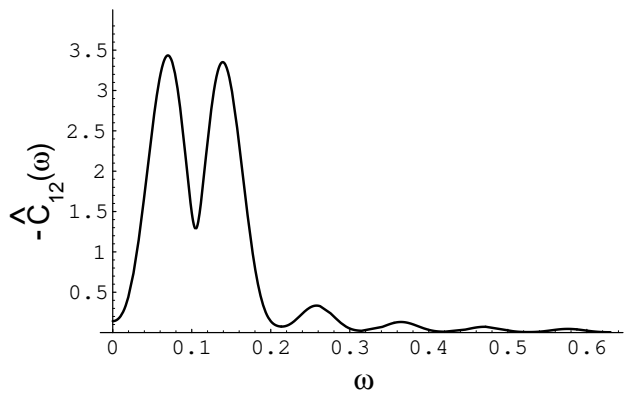
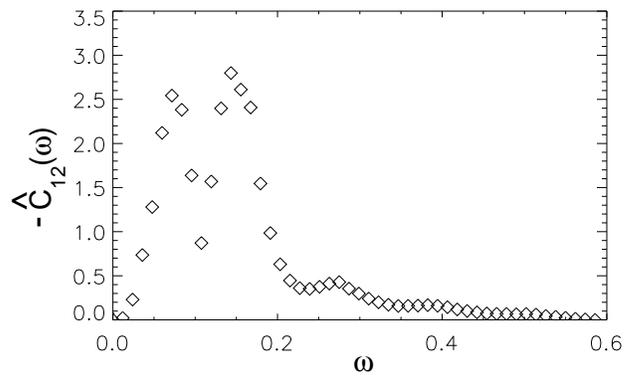
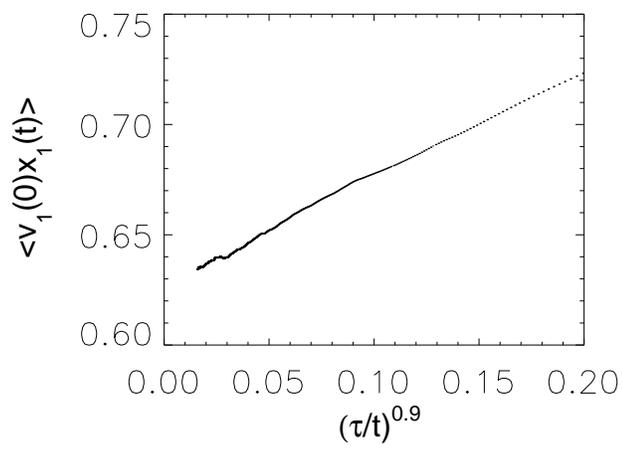
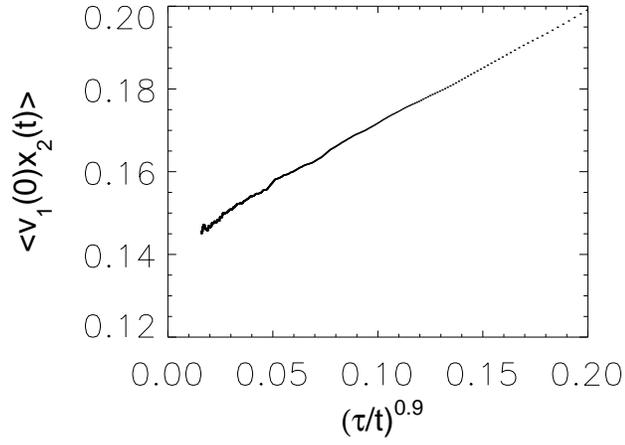
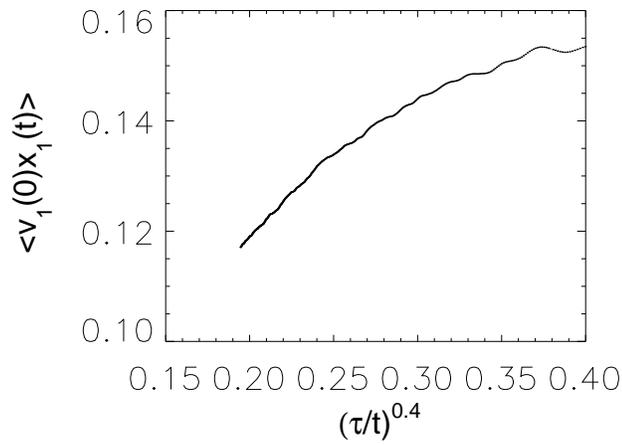
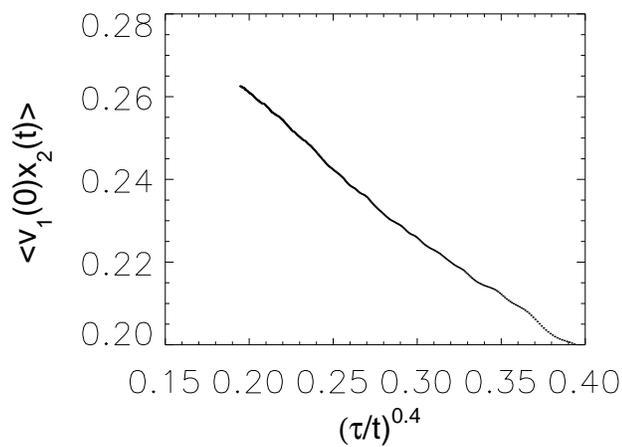
\end{document}